\documentclass{appolb}
\usepackage{amsmath}
\usepackage{amssymb,amsthm}
\usepackage{graphicx}
\begin{document}

\author{P.\ F.\ G\'ora\thanks{e-mail:gora@if.uj.edu.pl}
\address{M.\ Smoluchowski Institute of Physics and Complex Systems Research Center\\
Jagellonian University, Reymonta~4, 30-059 Krak\'ow, Poland}}

\title{Stationary distributions of a noisy logistic process}

\date{}

\maketitle

\begin{abstract}
Stationary solutions to a Fokker-Planck equation corresponding to a~noisy
logistic equation with correlated Gaussian white noises are constructed. Stationary
distributions exist even if the corresponding deterministic system
displays an unlimited growth. Positive correlations between the noises
can lead to a minimum of the variance of the process and to the stochastic
resonance if the system is additionally driven by a periodic signal.
\end{abstract}


\section{Introduction}

The logistic equation

\begin{equation}
\dot x = ax(1-x)\,,\ a>0\,,\ x\geqslant0\,,
\end{equation}

\noindent is one of the best-known and most popular models in population
dynamics. Perturbing this equation by a multiplicative noise is an obvious
generalization of the deterministic theory, aiming at describing populations
that live in an ever-changing environment. The logistic equation with 
a~fluctuating growth rate 

\begin{equation}\label{fp:logistic-noise-basic}
\dot x = (a+p\,\xi(t))x(1-x)\,,
\end{equation}

\noindent has been first discussed
by Leung in Ref.~\cite{Leung} and later by many other authors. Recently in 
Ref.~\cite{appb} we have discussed a further generalization of 
\eqref{fp:logistic-noise-basic} in which both the growth rate and the limiting population
level fluctuate, and these fluctuations are correlated in time:

\begin{equation}\label{fp:logistic-noise-general2}
\dot x = (a+p\,\xi_{\text{m}}(t))x - (b + q\,\xi_{\text{a}}(t))x^2\,.
\end{equation}

\noindent Here $\xi_{\text{m},\text{a}}$ are two Gaussian white noises (GWNs) that satisfy
$\left\langle\xi_i(t)\right\rangle=0$, $\left\langle\xi_i(t_1)\xi_i(t_2)\right\rangle=
\delta(t_1-t_2)$, $i=\text{m},\text{a}$, $\left\langle\xi_{\text{m}}(t_1)\xi_{\text{a}}(t_2)\right\rangle=
c\,\delta(t_1-t_2)$, $p$, $q$ are the amplitudes of the two noises  
and the correlation coefficient $c\in[-1,1]$. For the sake of terminology, we will sometimes
call the noise $\xi_{\text{m}}(t)$ ``multiplicative'' and the noise $\xi_{\text{a}}(t)$ ``additive'';
see Eq.~\eqref{fp:linear} below for a~rationale behind these names. Please note,
though, that on the level of Eq.~\eqref{fp:logistic-noise-general2} both these noises
are coupled multiplicatively to the process $x(t)$. Note also that
if $b>0$, the corresponding deterministic equation converges to a stable fixed point;
accordingly, we will call a system with a positive $b$ ``convergent''. If $b\leqslant0$,
the corresponding deterministic system displays unlimited growth. We will call a system with 
a~negative $b$ ``exploding''. Recently Mao \etal have shown in Ref.~\cite{marion} that in
the absence of $\xi_m$, the system \eqref{fp:logistic-noise-basic} remains positive
and bounded even in the ``exploding'' case. In  Ref.~\cite{njp} we have discussed certain
difficulties that may arise in numerical simulations of such a~system. This 
paper generalizes the result of Mao \etal to the case of both noises present.

In Ref.~\cite{appb} we have shown how the dynamics \eqref{fp:logistic-noise-general2} 
is related to the problem of 
a linear stochastic resonance. We have mapped the nonlinear equation
\eqref{fp:logistic-noise-general2} into a linear Langevin equation with two correlated
noises and used the solutions of the latter to heuristically explain the behaviour
of the noisy logistic equation. Specifically, the substitution

\begin{equation}\label{fp:substitution}
y=\frac{1}{x}
\end{equation}

\noindent converts Eq.~\eqref{fp:logistic-noise-general2} into a linear equation

\begin{equation}\label{fp:linear}
\dot y = -(a+p\,\xi_{\text{m}}(t))y + b + q\,\xi_{\text{a}}(t)
\end{equation}

\noindent which can be solved exactly for realizations of the process $y(t)$. 
The process \eqref{fp:linear} has a convergent mean if 
\begin{subequations}
\begin{equation}\label{fp:condition-mean}
a-\frac{1}{2}p^2>0
\end{equation}
and a convergent variance if a stronger condition
\begin{equation}\label{fp:condition-variance}
a-p^2>0
\end{equation}
\end{subequations}
holds. Using the properties of the process $y(t)$, useful prediction can be made 
about the noisy logistic process. Since in the presence of correlations, for certain 
values of parameters the variance of $y(t)$ first shrinks and then grows as a function
of the ``multiplicative'' noise strength, $q$, one expects a~similar behaviour for 
the logistic process $x(t)$ as well. These predictions have been corroborated
numerically in Ref.~\cite{appb}.
In particular, if $c=\pm1$, $bp\mp aq=0$ and the condition 
\eqref{fp:condition-variance} is satisfied, the variance of the process $y(t)$ vanishes.
The relation between the processes $y(t)$ and $x(t) = 1/y(t)$ \textit{intuitively}
means that if almost all realizations of the former asymptotically reach the same
constant value, so do almost all realizations of the latter. However, as
a~\textit{formal} relation between moments of these processes is not trivial, 
the predictions based on the properties of $y(t)$ have only a heuristic value. 

In the following we will
construct mathematically exact stationary solutions to the Fokker-Planck
equation corresponding to Eq.~\eqref{fp:logistic-noise-general2} and re-examine the above
results from the point of view of these stationary solutions. Furthermore, we
will show numerically that if the parameters undergo
periodic (for example, circaannual or seasonal) oscillations,
a positive correlation between the noises leads to a stochastic resonance.

In the Appendix we extend to the case of two correlated noises
the proof originally proposed by Mao \etal in Ref.~\cite{marion}
that solutions to Eq.~\eqref{fp:logistic-noise-general2}, when started from
a positive initial condition, never become negative almost surely.

\section{The Fokker-Planck equation}

The problem of constructing a Fokker-Planck equation corresponding to 
a~process driven by two correlated Gaussian white noises has been first
discussed in Ref.~\cite{Telejko}, where the two noises have been decomposed
into two independent processes. The same result has been later re-derived
in \cite{cao94}, where the authors have attempted to avoid an explicit
decomposition of the noises but eventually resorted to a disguised form of
the decomposition. The general Langevin equation

\begin{equation}\label{fp:general}
\dot x = h(x) + g_1(x)\xi_{\text{m}}(t) + g_2(x)\xi_{\text{a}}(t)\,,
\end{equation}

\noindent where $x(t)$ is a one-dimensional process and $\xi_{\text{m},\text{a}}$ are 
as in Eq.~\eqref{fp:logistic-noise-general2}, leads to the following Fokker-Planck 
equation in the Ito interpretation:

\begin{subequations}\label{fp:general-FP}
\begin{equation}
\frac{\partial P(x,t)}{\partial t} =
-\frac{\partial}{\partial x}h(x)P(x,t) +
\frac{1}{2}\frac{\partial^2}{\partial x^2}B(x)P(x,t)\,,
\end{equation}
\noindent where
\begin{equation}
B(x) = [g_1(x)]^2 + c\,g_1(x)g_2(x) + [g_2(x)]^2\,.
\end{equation}
\end{subequations}

In the case of Eq.~\eqref{fp:logistic-noise-general2} the corresponding Fokker-Planck
equation therefore reads

\begin{equation}\label{fp:FP}
\frac{\partial P(x,t)}{\partial t} =
-\frac{\partial}{\partial x}
\left[(a - bx)xP(x,t)\right]
+\frac{1}{2}\frac{\partial^2}{\partial x^2}
\left[x^2(p^2 - 2cpqx + q^2x^2)P(x,t)\right].
\end{equation}

\noindent It is apparent that the absolute signs of the two noise amplitudes
do not influence the solutions to the above equations, only their relative
sign, $\text{sgn}(pq)$, does. In the following we will assume that $\text{sgn}(pq)=+1$.
This comes at no loss to the generality as Eq.~\eqref{fp:FP} is invariant
under a simultaneous change of signs of $pq$ and the correlation coefficient,~$c$.

Stationary solutions to Eq.~\eqref{fp:FP} are the normalizable 
solutions to \cite{Risken,Gardiner}

\begin{equation}\label{fp:stationary}
x^2(p^2 - 2cpqx + q^2x^2)\frac{dP_{\text{st}}(x)}{dx}
+2x\left(2q^2x^2+(b{-}3cpq)x-(a{-}p^2)\right) P_{\text{st}}(x)
=0\,.
\end{equation}

A slight modification of the argument presented originally in Ref.~\cite{marion}
shows that all solutions to Eq.~\eqref{fp:logistic-noise-general2} that start
from a~positive initial condition remain positive almost surely; see the Appendix
for a~proof. Physically speaking, this results from a presence of an absorbing barrier
at $x=0$ in Eq.~\eqref{fp:logistic-noise-general2}: should the population suddenly
drop to zero, it would stay there forever. The formal result of Ref.~\cite{marion},
extended here to the case of two noises present, ensures that the
population never actually becomes nonpositive although in certain cases (see below)
it may dynamically cluster in a close proximity of $x=0^+$.
Therefore, we can divide both sides of \eqref{fp:stationary}
by $x$ and obtain

\begin{equation}\label{fp:stationary-x}
x(p^2 - 2cpqx + q^2x^2)\frac{d P_{\text{st}}(x)}{dx}
+2\left(2q^2x^2+(b-3cpq)x-(a-p^2)\right) P_{\text{st}}(x)
=0\,,
\end{equation}

\noindent provided $P_{\text{st}}(x)$ is normalizable over the $x>0$ semiaxix.

One may be tempted to try to immediately solve Eq.~\eqref{fp:stationary-x}
by standard methods, but a word of caution is needed here: 
should the coefficient
at $dP_{\text{st}}/dt$ vanish, special care must be taken.

\subsection{The case of only one noise present}

Before proceeding to the general case, we will discuss the special cases where 
only one of the amplitudes $p$, $q$ does not vanish.

\textit{Purely ``multiplicative'' noise.}
If $q=0$, the Langevin equation \eqref{fp:logistic-noise-general2} reduces to

\begin{equation}\label{fp:logistic-noise-reduced1}
\dot x = (a+p\,\xi_{\text{m}}(t))x - bx^2
\end{equation}

\noindent and we find for $P_{\text{st}}(x)$ 

\begin{equation}\label{fp:special1}
P_{\text{st}}(x)
={\cal N} x^{2(a-p^2)/p^2} \exp\left(-\frac{2bx}{p^2}\right),
\end{equation}

\noindent where $\cal N$ is a normalization constant. This function is normalizable
for $x>0$ if $b>0$, or if the system is convergent, and

\begin{subequations}
\begin{equation}\label{fp:condition1}
a-\frac{1}{2}p^2>0\,,
\end{equation}
which determines the behaviour of the distribution around $x=0$. Moreover, if

\begin{equation}\label{fp:condition2}
a-p^2>0\,,
\end{equation}
\end{subequations}

\noindent
$P_{\text{st}}(x)$ goes to zero as $x\to0^+$ and has a maximum at $x=(a-p^2)/b$.
Note that the conditions \eqref{fp:condition1}, \eqref{fp:condition2} coincide
with the conditions \eqref{fp:condition-mean}, \eqref{fp:condition-variance}
determining the properties of the linear process \eqref{fp:linear}. For 
$p^2> a>\frac{1}{2}p^2$, the distribution is mildly divergent at $x=0$ and decreases 
monotically with an increasing $x$.

\textit{Purely ``additive'' noise.}
If $p=0$, the Langevin equation takes the form

\begin{equation}\label{fp:logistic-noise-reduced2}
\dot x = (a - bx)x + qx^2\xi_{\text{a}}(t)
\end{equation}

\noindent and we obtain for the stationary distribution

\begin{equation}\label{fp:special2}
P_{\text{st}}(x) = \frac{\cal N}{x^4}\exp\left(-\frac{a-2bx}{q^2x^2}\right)
\end{equation}

\noindent which is normalizable whenever $a>0$. Note that no bounds on $b$ are 
imposed: The stationary distribution exists for both convergent and exploding systems.
This special case 
falls into a broad category discussed recently by Mao \etal in Ref.~\cite{marion},
without further generalizations provided by the present paper.
The distribution \eqref{fp:special2} has a maximum at the positive root of

\begin{equation}\label{fp:special2-minimum}
2q^2x^2 + bx -a=0\,.
\end{equation}

\subsection{The general case}

If both noises are present and are not maximally correlated, $|c|\not=1$,
the solution to \eqref{fp:stationary-x} reads

\begin{equation}\label{fp:solution-general}
P_{\text{st}}(x) = 
\frac{{\cal N}x^{2(a-p^2)/p^2}}{(p^2{-}2cpqx {+} q^2x^2)^{(a+p^2)/p^2}}
\exp\left[
-\frac{2(bp{-}acq)\arctan\left(\frac{qx-cp}{\sqrt{1 - c^2}p}\right)}{\sqrt{1-c^2}\,p^2q}
\right].
\end{equation}

\begin{figure}
\begin{center}
\includegraphics{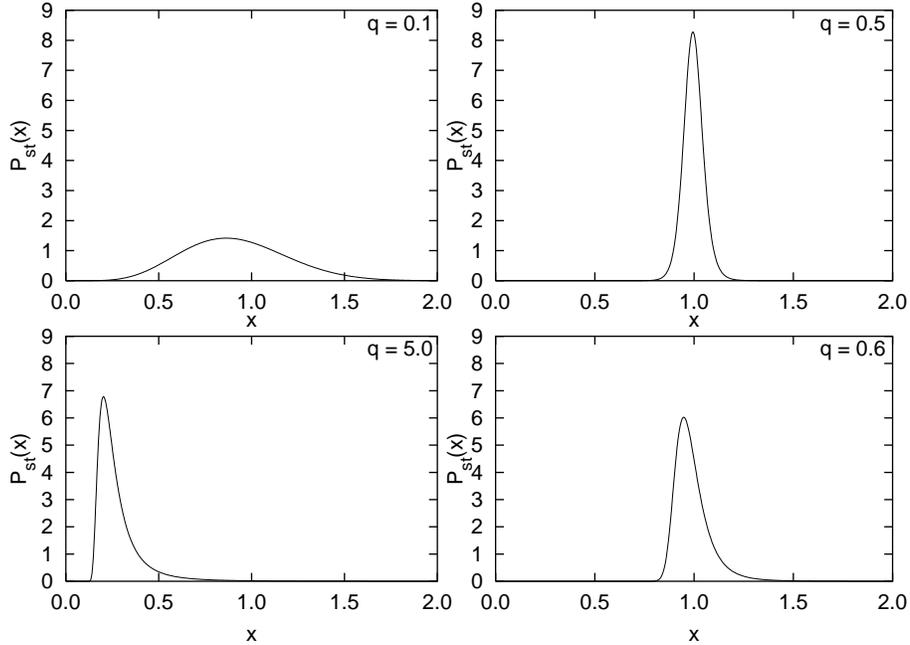}
\end{center}
\caption{Stationary distributions \eqref{fp:solution-general} in a strongly
correlated case, $c=0.99$. Clockwise, from top-left $q=0.1$, $q=0.5$, $q=0.6$,
and $q=5.0$. Other parameters, common for all panels, are $p=0.5$, $a=b=1$.}
\label{fp:fig1}
\end{figure}

\begin{figure}
\begin{center}
\includegraphics{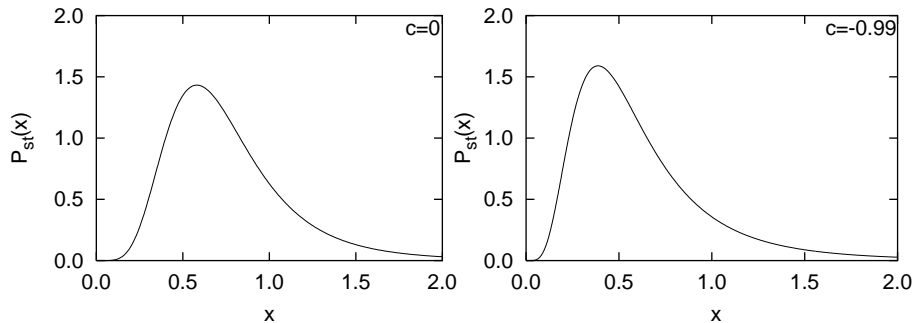}
\end{center}
\caption{Stationary distributions for the uncorrelated ($c=0$, left panel) and
a~strongly anticorrelated ($c=-0.99$, right panel) cases. Other parameters are
$a=b=1$, $p=q=0.5$.}
\label{fp:fig2}
\end{figure}

\noindent Since the exponential term is limited, the convergence (normalization)
properties of \eqref{fp:solution-general} are determined by those of the fractional
term. The denominator is always strictly positive.
For $x\to\infty$, $P_{\text{st}}(x) \sim x^{-4}$
for all possible values of parameters. $P_{\text{st}}(x)$ is therefore
normalizable if it does not diverge too rapidly at $x\to0^+$, or again if the
condition \eqref{fp:condition1} holds. Either in this case, no bounds on $b$ are imposed. 
If the condition \eqref{fp:condition2} holds
as well, the distribution \eqref{fp:solution-general} approaches zero as $x\to0^+$,
but this condition is no longer associated with the presence of a maximum. 
The maximum of $P_{\text{st}}(x)$ coincides with the
positive root of

\begin{equation}\label{fp:root}
2q^2x^2 + (b-3cpq)x -(a-p^2) =0\,,
\end{equation}

\noindent cf.~Eq.~\eqref{fp:stationary-x}, provided such a root exists. 
It certainly does for $a-p^2>0$, but it can appear also for
$p^2>a>\frac{1}{2}p^2$, where the distribution is mildly divergent.
Example stationary distributions in a strongly correlated case are presented
on Fig.~\ref{fp:fig1}. For large values of the additive noise strength, $q$,
the distributions are highly skewed and squeezed against the $x=0$ axis.
For comparison, on Fig.~\ref{fp:fig2} we show example stationary distributions 
for the uncorrelated and a strongly anticorrelated cases. The distributions
presented are skewed and much wider than the distribution from Fig.~\ref{fp:fig1}
with the same value of $q=0.5$. These distributions
also get squeezed as the additive noise strength becomes large.
Note that the distribution corresponding to the anticorrelated noises is already
more squeezed than the distribution for the uncorrelated case.

If the distribution \eqref{fp:solution-general} is normalizable, it has a~convergent
mean and a~variance. Its higher moments are divergent.

\subsection{The maximally correlated case}

If the two noises are maximally (anti)correlated, $c=\pm1$, 
Eq.~\eqref{fp:stationary-x} takes the form

\begin{equation}\label{fp:stationary-resonant}
x(p\mp qx)^2\frac{dP_{\text{st}}}{dx} 
+ 2\left[(p\mp qx)^2 + q^2x^2 +(b\mp pq)x -a\right]P_{\text{st}} = 0\,,
\end{equation}

\noindent leading to the following candidate solution:

\begin{equation}\label{fp:stationary-resonant-solution}
P_{\text{st}}(x) = {\cal N}
\frac{x^{2(a-p^2)/p^2}}{(p\mp qx)^{2(a+p^2)/p^2}}
\exp\left[\mp\frac{2(bp\mp aq)}{pq(p\mp qx)}\right].
\end{equation}

With our sign convention adopted, $\text{sgn}(pq)=+1$, this solution
is normalizable if $c=-1$ and the condition 
\eqref{fp:condition1} holds. The distribution \eqref{fp:stationary-resonant-solution} 
with the ``$+$'' sign can be obtained from 
Eq.~\eqref{fp:solution-general} by taking the limit $c\to-1$. 
This distribution decreases as $x^{-4}$ with $x\to\infty$. 
The maximum
of \eqref{fp:stationary-resonant-solution}, if it exists, coincides with the
positive root of Eq.~\eqref{fp:root} with $c=-1$.

The case of $c=+1$ is more challenging. First, if the resonant condition

\begin{equation}\label{fp:resonantcondition}
bp-aq=0
\end{equation}

\noindent holds, Eq.~\eqref{fp:stationary-resonant} is solved by

\begin{equation}\label{fp:delta}
P_{\text{st}}(x) = \delta\left(x-\frac{p}{q}\right)
\end{equation}

\noindent regardless of the value of $p$.
This result is stronger than that reported in Ref.~\cite{appb}
where we could predict a $\delta$-shaped distribution only in the $a\nobreak-\nobreak{}p^2>0$ case,
or when the variance of the corresponding linear system was convergent, as otherwise
any predictions based on the linear system failed. Note that with the condition
\eqref{fp:resonantcondition} statisfied, Eq.~\eqref{fp:logistic-noise-general2}
reduces to a rescaled form of Eq.~\eqref{fp:logistic-noise-basic}.

If $c=+1$ and the condition \eqref{fp:resonantcondition} does not hold, 
Eq.~\eqref{fp:stationary-resonant} does not have a~normalizable solution.
This observation is slightly surprising, but formally speaking, it results
from the fact that the double limit

\begin{equation}\label{fp:doublelimit}
\lim\limits_{\begin{array}{c}\scriptstyle x\to p/q\\\scriptstyle c\to1\end{array}}\!\!
\exp\left[
-\frac{2(bp-acq)\arctan\left(\frac{qx-cp}{\sqrt{1 - c^2}p}\right)}{\sqrt{1-c^2}\,p^2q}
\right]
\end{equation}

\noindent does not exits: Its value depends on which route the singularity is approached.
The nonexistence of stationary solutions in the fully correlated, non-resonant
case is, therefore, related to the essential singularity of the complex exponential
at infinity. The fact that with $c=+1$, $bp-aq\not=0$, the drift and diffusive terms in
Eq.~\eqref{fp:stationary-resonant} both vanish, but at different points, is the physical
reason for this apparent oddity: If the population gets located around the point of the
vanishing diffusion, it is washed away by the drift, and if it gets located around
the point of the vanishing drift, it diffusively leaks from there.
Nevertheless, if $a-\frac{1}{2}p^2>0$, in numerical simulations the cases of
$c=1$ and $c=1-\varepsilon$ with $0<\varepsilon\ll1$ are undistinguishable. In the
latter case, the distribution \eqref{fp:solution-general} is perfectly
normalizable.

\section{Resonant effects and the shape of the stationary distribution}

Perhaps the most important prediction based on the analysis of the linear equation
\eqref{fp:linear} and discussed in Ref.~\cite{appb} is that, for certain values of parameters,
the variance of the process $x(t)$ should, in the asymptotic regime, first shrink,
reach a minimum, and then grow as a function of the additive noise strength, $q$.
As we have mentioned before, these are heuristic, intuitive conclusions based on
the behaviour of the linear system associated with the logistic process, but because
of the complicated relation between the moments of these two processes, they do not
amount to a formal proof. In Ref.~\cite{appb} we have confirmed these predictions 
numerically for a certain range of the additive noise strengths. As we have seen
above, in the fully correlated and resonant case, 
the stationary distribution becomes $\delta$-shaped and its variance indeed vanishes,
much as predicted by the linear system. Since we now know the mathematically exact
stationary distributions, we can test the behaviour of the variance in the general case
directly.

\begin{figure}
\begin{center}
\includegraphics{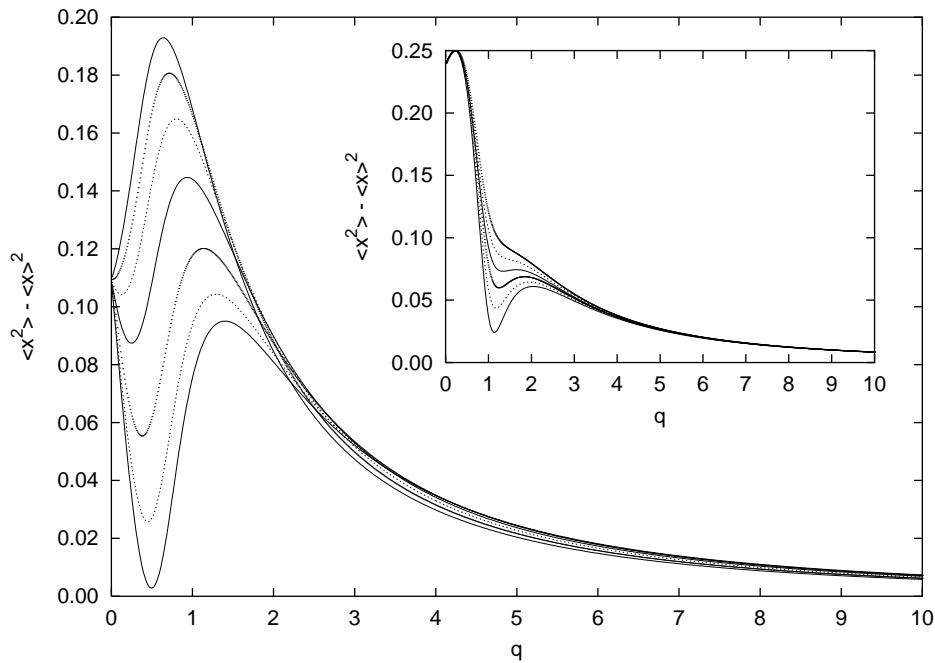}
\end{center}
\caption{The variance $\left\langle x^2\right\rangle-\left\langle x\right\rangle^2$
determined form the distribution \eqref{fp:solution-general} as a function of the
additive noise strength, $q$. Main panel: $p=0.5$, the curves, from bottom to top,
correspond to $c=0.99$, $c=0.90$, $c=0.75$, $c=0.50$, $c=0.25$, $c=0$, and $c=-0.25$,
respectively. Inset: $p=1.1$, the curves correspond, from bottom to top,
to $c=0.99$, $c=0.98$, $c=0.97$, $c=0.96$, $c=0.95$, and $c=0.94$, respectively.
Other parameters, common for all curves presented, are $a=b=1$.}
\label{fp:fig3}
\end{figure}

Recall that the distribution \eqref{fp:solution-general} has the two first moments
convergent whenever it is normalizable. Unfortunately, analytical expressions for
these moments cannot be obtained, mainly due to the presence of the complicated
exponential term. Therefore, we have calculated the moments by numerically integrating
over the distribution \eqref{fp:solution-general}. Results are presented on 
Fig.~\ref{fp:fig3}. If the distribution approaches zero as $x\to0^+$, or when the
condition \eqref{fp:condition2} is satisfied, and if the two noises are positively
correlated, $0<c<1$, the variance $\left\langle x^2\right\rangle-
\left\langle x\right\rangle^2$ displays a clear minimum as a function of the 
additive noise strength, $q$. The minimum becomes shallower as the correlations
decrease towards zero, where it eventually disappears. It is not present for the
negative correlations or when one of the amplitudes vanishes. These effects agree 
with predictions based on the linear system \eqref{fp:linear}. The presence of 
the minimum of the variance is a clear and beneficial effect of positive correlations 
between the two noises. However, for larger values of $q$ a new phenomenon appears:
The variance starts decreasing again. This is because for large values of the additive noise,
the stationary distribution
gets squeezed against the $x=0$ axis, cf.~Fig.~\ref{fp:fig1} above. This effect
cannot be predicted within the linear approach --- note that the process
described by Eq.~\eqref{fp:linear} has a support that formally spreads
over the entire real axis and, moreover, is Gaussian whenever the condition
\eqref{fp:condition2} holds, while the noisy logistic process is restricted to the
positive semiaxix.

If the stationary distribution mildly diverges at zero, or if $p^2>a>\frac{1}{2}p^2$,
a distinct minimum in the variance of $x$ also appears but it is present only for
fairly large (and positive) values of the correlation coefficient, 
cf.\ the inset on Fig.~\ref{fp:fig3}.
Note that this effect cannot possibly be predicted by analysing the linear system
\eqref{fp:linear} as in this regime the variance of the linear process diverges
and any predictions break.

\begin{figure}
\begin{center}
\includegraphics{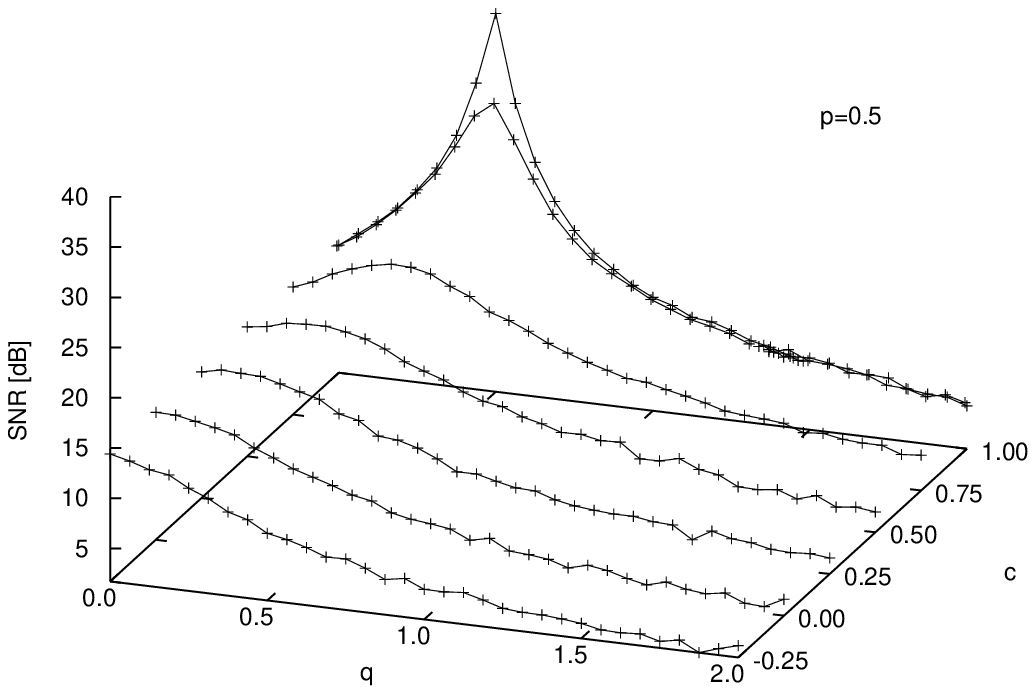}
\includegraphics{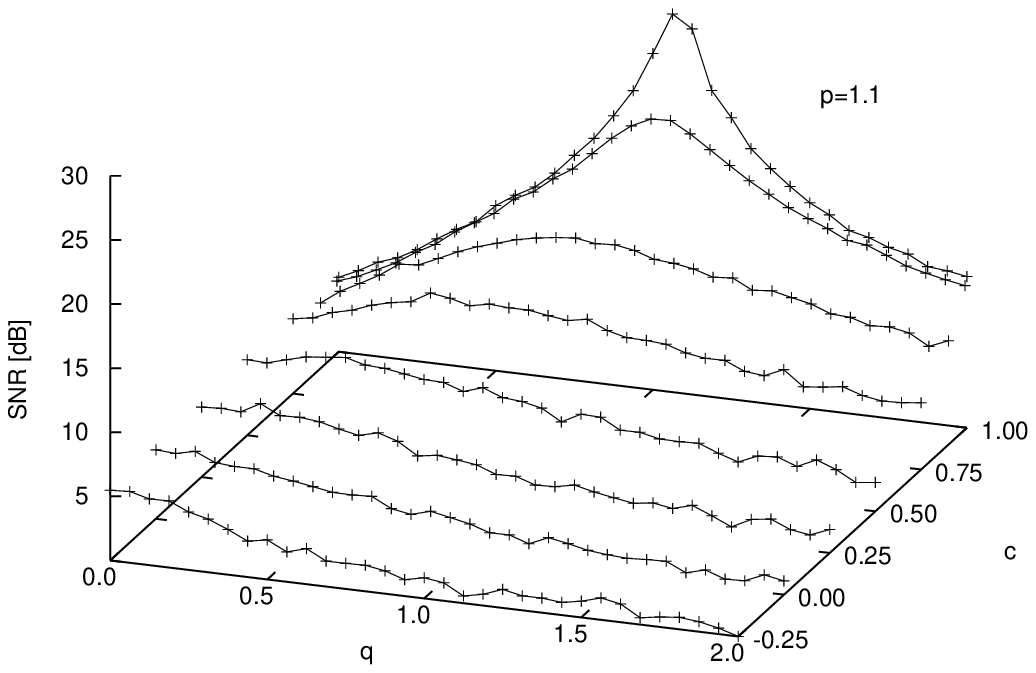} 
\end{center}
\caption{Stochastic resonance in the system \eqref{fp:logistic-noise-oscillations}.
The upper panel --- the condition \eqref{fp:condition2} is satisfied,
$p=0.5$. The lower panel --- the condition \eqref{fp:condition2} is not satisfied, 
$p=1.1$. Other parameters, common for the two panels, are $a=b=1$, $A=0.5$,
$\Omega=2\pi$. Curves presented correspond, back to front, to $c=1.0$, $0.99$, $0.9$ (lower panel only),
$0.75$, $0.5$, $0.25$, $0.0$, and $-0.25$, respectively.}\label{fp:fig4}
\end{figure}

\section{Stochastic resonance}

We now assume that parameters of the logistic process are not only subjected to noise,
but also to periodic, deterministic perturbations, resulting for example from seasonal
changes in the environment. Specifically, we consider

\begin{equation}\label{fp:logistic-noise-oscillations}
\dot x = (a+p\,\xi_{\text{m}}(t))x - (b+A\sin(\Omega t +\varphi) + q\,\xi_{\text{a}}(t))x^2\,.
\end{equation}

We have shown analytically in Ref.~\cite{appb} that the linear system associated
with Eq.~\eqref{fp:logistic-noise-oscillations} displays a stochastic resonance (SR) if the
noises are positively correlated. SR is one of the most spectacular examples of 
a~constructive role of noise --- see Ref.~\cite{SRreview} for a review.
Because we do not know exact solutions of 
a~time-dependent Fokker-Planck equation corresponding to 
Eq.~\eqref{fp:logistic-noise-oscillations}, we will demonstrate the SR phenomenon
numerically. We will use the Signal-To-Noise Ratio (SNR) as a~measure of the SR:

\begin{equation}\label{appb:SNR-def}
\mathrm{SNR} = 
10 \log_{10}\frac{P_{\text{signal}}}{P_{\text{noise}}(\omega=\Omega)}\,,
\end{equation}
\noindent where $P_{\text{signal}}$ is the height of the peak in the 
power spectrum at the driving frequency and $P_{\text{noise}}$ is the noise-induced
background.

We have solved the equation \eqref{fp:logistic-noise-oscillations} numerically
with the Euler-Maryuama algorithm and a timestep equal $2^{-16}$. The GWNs have
been generated by the Marsaglia algorithm \cite{Marsaglia} and the famous
Mersenne Twister \cite{Mersenne} has been used as the underlying uniform generator.
We have let the system to equilibrate, run the simulations
for $2^{25}$ steps and collected the results of every $2^9$-th step, 
calculated the power spectrum, calculated the 
SNR and averaged the results over 128 realizations of the stochastic
processes and the initial phases,~$\varphi$. Selected results are presented on
Fig.~\ref{fp:fig4}.

The upper panel corresponds to the situation when the condition \eqref{fp:condition2}
holds, or when the noisy logistic process without the periodic signal has a~convergent
variance. We can clearly see that the system \eqref{fp:logistic-noise-oscillations}
displays a~SR for positive correlations between the two noises and disappears
for $c\leqslant0$: For positive correlations between the noises, there is a certain
level of the ``additive'' noise that maximizes the impact that seasonal
changes in the environment have on the population. As we have 
shown in the preceding Section, this range of parameters corresponds to the presence
of the minimum in the variance of \eqref{fp:solution-general}. One may be tempted 
to conclude that the SR and the minimum of the variance are two facets of the
same phenomenon, much as in the linear case. However, the lower panel of Fig.~\ref{fp:fig4},
corresponding to the situation when the condition \eqref{fp:condition2} is not
satisfied, shows that this is not the case. The SR, albeit much weaker than in the
previous case, is clearly present even when the signal-free system no longer displays 
a minimum of the variance. A minimum of the variance and the stochastic resonance 
are two different constructive effects of positive correlations between the noises.

\section{Conclusions}

In this paper we have constructed stationary distributions corresponding to a noisy
logistic process driven by two correlated GWNs. These distributions are restricted
to the positive semiaxis and if they are normalizable, they have two (and only two)
convergent moments. In particular, if the noises are maximally correlated and a certain
resonant condition holds, the stationary distribution is $\delta$-shaped, which
has been reported previously as a result of many numerical simulations. Surprisingly,
if the noises are maximally correlated but the resonant condition does not hold,
the process does not have a stationary distribution,
even though it can numerically manifest itself as if it had one.

Positive correlations between the noises lead to a minimum of the variance of the
noisy logistic process and to a stochastic resonance if the parameters of the 
system undergo additional periodic changes. As we have numerically demonstrated,
these are two different effects. By constructing the exact stationary distributions,
we have extended our previous analysis of the system performed mainly by formally 
converting the system into a~linear one. Several features of the system,
and ``squeezing'' of the stationary distribution in case of
a strong ``additive'' noise in particular, cannot be described by analysing the 
linear process. This
is because the linear process \eqref{fp:linear} is Gaussian if it has two convergent
moments but the nonlinear logistic process is not.
Nevertheless, there are nice parallels between the 
properties of the logistic process and its formal linearization: 
If $a-\frac{1}{2}p^2>0$, the linear process has a convergent mean and the
logistic process has a normalizable stationary distribution that decreases for large 
$x$ as $x^{-4}$. If $a-p^2>0$, the stationary distribution of the logistic process approaches
zero as $x\to0^+$ and the linear process has a convergent variance.
If $\frac{1}{2}p^2<a<p^2$, the stationary distribution of the noisy logistic
process mildly diverges at $x=0^+$ and the population dynamically clusters around that
point. Note that for $a>p^2$, the stationary distribution has a maximum and the population
is actually pushed away from $x=0^+$. Thus, the level of the multiplicative noise
$p^2=a$ marks a qualitative change in the population described by the noisy logistic
equation.

It is, perhaps, surprising that a stationary distribution of the noisy logistic
process may exist even if the corresponding deterministic process is clearly
divergent. The fact that noise can prevent a population from
exploding has been recently reported by Mao \etal in Ref.~\cite{marion} for 
a~more restricted, in a sense, class of systems. The present work is
an extension of this research to a~class that includes two correlated
sources of the noise.

I am very grateful to Prof.\ Andrzej Fuli\'nski for his thorough criticism of 
a~draft version of this paper.

\appendix

\section{}

To show that solutions to Eq.~\eqref{fp:logistic-noise-general2} remain positive
almost surely when started from a positive initial condition, we first
decompose the correlated noises  $\xi_{\text{m}}$, $\xi_{\text{a}}$ into
two independent processes:

\begin{subequations}\label{fp:decomposition}
\begin{eqnarray}
\xi_{\text{m}}(t) &=& \xi(t)\,,\\
\xi_{\text{a}}(t) &=& c\xi(t) + \sqrt{1-c^2}\,\eta(t)
\end{eqnarray}
\end{subequations}

\noindent where $\eta(t)$, $\xi(t)$ are two identical, uncorrelated GWNs.
The decomposition \eqref{fp:decomposition} is a variant of the method originally
used in Ref.~\cite{Telejko}. 
We now cast Eq.~\eqref{fp:logistic-noise-general2} in a form customarily used by
mathematicians:

\begin{equation}\label{fp:wiener}
dx = (ax-bx^2)\,dt + (px-qcx^2)\,du -q\sqrt{1-c^2}\,x^2\,dw\,,
\end{equation}

\noindent where $du$, $dw$ are differentials over two identical, independent 
Wiener processes. Incidentally, observe that the Fokker-Planck equation \eqref{fp:FP}
follows immediately from Eq.~\eqref{fp:wiener}.

\newtheorem{positive}{Theorem}

\begin{positive}\label{fp:theorem}
If the initial condition $x_0>0$, for any $q\not=0$ the solution to 
Eq.~\eqref{fp:wiener} remains positive for all $t>0$ almost surely.
\end{positive}

\begin{proof}

The proof of this theorem follows closely that of Theorem~2.1 from the work of Mao
\etal \cite{marion} and we encourage readers interested in mathematical details to 
familiarize with that proof first; to save the space, we will show only this
part in which the proof of Theorem~\ref{fp:theorem} differs from that of Mao \etal

First, the authors of Ref.~\cite{marion} consider a multispecies (multidimensional)
system, while we restrict ourselves to a simpler single-species case.

Second, the proof is based on properties of the function 

\begin{equation}\label{fp:V}
V(s) = \sqrt{s} - 1 - {\textstyle\frac{1}{2}}\ln s\,.
\end{equation}

\noindent This function is nonnegative for any $s>0$. We calculate $V(x(t))$ along
the trajectory generated by Eq.~\eqref{fp:wiener} with an initial condition $x_0>0$
and calculate the stochastic differential of $V(x(t))$ using Ito formula. 
Because Eq.~\eqref{fp:wiener} differs from that considered by Mao \etal, we obtain 
a slightly different expression. Specifically, if $x(t)>0$,

\begin{gather}
dV(x(t)) = {\textstyle\frac{1}{2}}\left(x^{-1/2}-x^{-1}\right)
\!
\left[(ax{-}bx^2)\,dt + (px{-}qcx^2)\,du -q\sqrt{1{-}c^2}x^2\,dw\right]
\nonumber\\
\label{fp:dV}
{}+
{\textstyle\frac{1}{4}}\left(x^{-2}-{\textstyle\frac{1}{2}}x^{-3/2}\right)
\left[(px-qcx^2)^2+q^2(1-c^2)x^4\right]\,dt
\end{gather}

\noindent The second term in \eqref{fp:dV} would be absent if the noises were
interpreted in the Stratonovich sense. After a simple algebra,

\begin{eqnarray}
dV(x(t))&=&\left[{\textstyle\frac{1}{2}}\left(x^{1/2}-1\right)(a{-}bx)
+{\textstyle\frac{1}{4}}\left(1-{\textstyle\frac{1}{2}}x^{1/2}\right)(p^2{-}2pqcx{+}q^2x^2)\right]
dt
\nonumber\\\label{fp:dV2}
&-&{\textstyle\frac{1}{2}}\left(x^{1/2}-1\right)(p-qcx)\,du 
-{\textstyle\frac{1}{2}}\left(x^{1/2}-1\right)q\sqrt{1-c^2}\,x\,dw\,.
\nonumber\\
\end{eqnarray}

If $q=0$ and $b<0$, or when the corresponding deterministic system explodes, the
coefficient at $dt$ in \eqref{fp:dV2} may assume arbitrarily large values.
On the contrary, for any $q\not=0$ and regardless of the sign of $b$,
this coefficient is bounded from above by a certain positive number $K$. 
Thus\footnote{Some subtleties of the notation are omitted here, see~\cite{marion} 
for a fully rigorous treatment.}

\begin{eqnarray}\label{fp:dV-calka}
\int_0^T dV(x(t)) &\leqslant& KT 
-\int_0^T {\textstyle\frac{1}{2}}\left(x^{1/2}-1\right)(p-qcx)\,du 
\nonumber\\
&&\hphantom{KT}
-\int_0^T {\textstyle\frac{1}{2}}\left(x^{1/2}-1\right)q\sqrt{1-c^2}\,x\,dw\,,
\end{eqnarray}

\noindent where $T$ is a time such that $x(t)$ is positive for
$0\leqslant t<T$ almost surely. By taking the expectation values, we obtain

\begin{equation}\label{fp:V(t)}
\left\langle V(x(T))\right\rangle \leqslant V(x_0) + KT\,.
\end{equation}

The rest of the proof now proceeds exactly as in Ref.~\cite{marion} to
show that $T=\infty$.
\end{proof}


\begin{thebibliography}{10}


\bibitem{Leung} H. K. Leung, Phys. Rev. A {\bf37}, 1341 (1988).

\bibitem{appb} P. F. G\'ora, Acta Phys. Pol. B {\bf35}, 1583 (2004) .

\bibitem{marion} X. Mao, G. Marion, and E. Renshaw, Stochastic Process. Appl. 
{\bf 97}, 95 (2002).

\bibitem{njp} P. F. G\'ora, New J. Phys. {\bf7}, 36 (2005). 

\bibitem{Telejko} A. Fuli\'nski and T. Telejko, Phys. Lett. A {\bf 152} (1991) 11.

\bibitem{cao94} Wu Da-jin, Cao Li, and Ke Sheng-zi, Phys. Rev. E {\bf50}, 2496 (1994).

\bibitem{Risken} H. Risken, \textit{The Fokker-Planck Equation} 
(Springer, Berlin, 1984).

\bibitem{Gardiner} C. W. Gardiner, \textit{Handbook of Stochastic Methods} 
(Springer, Berlin, 1993).

\bibitem{SRreview} L. Gammaitoni, P. H\"anggi, P. Jung, and F. Marchesoni, 
Rev. Mod. Phys. {\bf70}, 223 (1998).

\bibitem{Marsaglia} G. Marsaglia and T. A. Bray, SIAM Review {\bf6}, 260 (1964);
A. J. Kinderman and J. G. Ramage, J. Amer. Statist. Assoc. {\bf 71}, 893 (1976);
R. Wieczorkowski and R. Zieli\'nski, \textit{Komputerowe generatory liczb losowych}
(WNT, Warszawa, 1997) (in Polish).

\bibitem{Mersenne} M. Matsumoto and T. Nishimura, ACM Trans. on Modeling and 
Computer Simulation, {\bf 8}, 3 (1998).

\end{thebibliography}
\end{document}